\documentclass[journal]{IEEEtran}
\usepackage{graphicx}
\usepackage{amsmath}
\usepackage{amsfonts}
\usepackage{algorithm}
\usepackage{algorithmic}
\usepackage{makecell}
\usepackage{booktabs}
\usepackage{multirow}
\usepackage{subfigure} 
\usepackage{color}
\usepackage{comment}

\setcellgapes{3pt}
\setcellgapes{3pt}

\usepackage[justification=centering]{caption}

\usepackage{cite}

\ifCLASSINFOpdf
\else
\fi
\begin{document}
%
\title{AI Driven Heterogeneous MEC System with UAV Assistance for Dynamic Environment - Challenges and Solutions}

\author{Feibo Jiang, Kezhi Wang, Li Dong, Cunhua Pan, Wei Xu, and Kun Yang.
	\thanks{	
		Feibo Jiang  (jiangfb@hunnu.edu.cn) is with Hunan Provincial Key Laboratory of Intelligent Computing and Language Information Processing, Hunan Normal University, Changsha, China, Kezhi Wang (kezhi.wang@northumbria.ac.uk) is with the department of Computer and Information Sciences, Northumbria University, UK, Li Dong (Dlj2017@hunnu.edu.cn) is with Key Laboratory of Hunan Province for New Retail Virtual Reality Technology, Hunan University of Commerce, Changsha, China, Cunhua Pan (Email: c.pan@qmul.ac.uk) is with School of Electronic Engineering and Computer Science, Queen Mary University of London, London, E1 4NS, UK, Wei Xu (wxu@seu.edu.cn) is with NCRL, Southeast University, Nanjing, China, Kun Yang (kunyang@essex.ac.uk) is with the School of Computer Sciences and Electrical Engineering, University of Essex, CO4 3SQ, Colchester, UK and also with University of Electronic Science and Technology of China, Chengdu, China
	
				(Corresponding authors: Kezhi Wang; Li Dong.)
			}
}

\markboth{IEEE Network Magazine}%
{Shell \MakeLowercase{\textit{et al.}}: Bare Demo of IEEEtran.cls for IEEE Journals}
%



\maketitle

\begin{abstract}

By taking full advantage of Computing, Communication and Caching (3C) resources at the network edge, Mobile Edge Computing (MEC) is envisioned as one of the key enablers for the next generation networks. However, current fixed-location MEC architecture may not be able to make real-time decision in 
dynamic environment, especially in large-scale scenarios. To address this issue, in this paper, a Heterogeneous MEC (H-MEC) architecture is proposed, which 
is composed of fixed unit, i.e., Ground Stations (GSs) as well as moving nodes, i.e., Ground Vehicles (GVs) and Unmanned Aerial Vehicles (UAVs), all with 3C resource enabled.
The key challenges in H-MEC, i.e., mobile edge node management, real-time decision making, user association and resource allocation along with the possible Artificial Intelligence (AI)-based solutions are discussed. 
In addition, the AI-based joint Resource schEduling (ARE) framework with two different AI-based mechanisms, i.e., Deep neural network (DNN)-based and deep reinforcement learning (DRL)-based architectures are proposed. DNN-based solution with online incremental learning applies the global optimizer and therefore has better performance than the DRL-based architecture with online policy updating, but requires longer training time. 
The simulation results are given to verify the efficiency of our proposed ARE framework.

\end{abstract}

\begin{IEEEkeywords}
Heterogeneous mobile edge computing, artificial intelligence, deep neural network, deep reinforcement learning, dynamic environment.
\end{IEEEkeywords}

%
\IEEEpeerreviewmaketitle

\section{Introduction}

%
%
%
%

Recently, with the increasing popularity of new resource-intensive applications, e.g., automatic driving, online gaming, health monitoring, and Virtual Reality (VR) services, the quality of our life is expected to be improved significantly. In addition, there is a growing trend to execute the above attractive applications in our User Equipments (UEs), e.g., mobile phones or handheld devices. However, this contradicts to the sizes and the battery capacities of the UEs. 

Fortunately, Mobile Edge Computing (MEC) \cite{8016573} has been proposed by taking full advantage of cooperation between Communication, Computation and Caching (3C) resources at the network edge. Specifically, MEC can enable UEs with computational-intensive tasks to offload them to the edge cloud and is envisioned as one of the key enabling technologies for the next generation mobile networks \cite{8754787}.

The traditional MEC system is not flexible and may suffer from high deployment cost, due to its fixed architecture. However, the future networks are expected not only to accommodate an unprecedented dynamic and heterogeneous environment, but also should be able to support on-demand hotspot areas and temporary activities, in a fast and highly reliable manner. In other words, in the future, we envision more flexible user patterns and services, i.e., the number, the locations and the service requirements of the mobile users may be constantly changing. Hence, the current MEC system cannot be applied in future networks due to its fixed architecture.

To address the above-mentioned problem, in this paper,
we propose a Heterogeneous MEC (H-MEC) system, which
is composed of fixed Ground Stations (GSs), mobile Unmanned Aerial Vehicles (UAVs) and Ground Vehicles (GVs), all equipped with 3C resources. The fixed MEC node, such as GS, can be charged from the power grid and cooled by the air conditioner, whereas the mobile nodes, such as GV and UAV can be charged from the charging pile on the roadside and roof respectively.
H-MEC is more flexible than the traditional MEC system and is more suitable in the dynamic environment, as UAVs and GSs can be deployed on demand. In addition, UAV and GV can move close to the user to improve the connectivity. However, various research challenges arise when applying H-MEC in dynamic environment, such as mobile edge node management, real-time decision making, fast user association and resource allocation.

 Fortunately, these challenges fall into the the research of artificial intelligence (AI), which is considered to be a promising technique to address such problems by adaptive modelling and intelligent learning. Different from the traditional optimization methods (e.g., convex optimization, dynamic programming and game theory), AI based solution has continuous learning ability for the dynamic environment and can make the real-time inference with low computational complexity.

In this paper, we summarize the key challenges in deploying H-MEC in dynamic environment and propose AI-based solutions to tackle these issues. The contributions of this paper are summarized as follows:

(1) We first discuss the AI driven H-MEC architecture and then summarize the typical applications of this architecture.

(2) Next, we show the key challenges of H-MEC applied in dynamic environment, i.e., in the scenarios where the number, the locations and the requirement of the UEs are constantly changing. Moreover, we show possible AI-based solutions to address the above challenges.

(3) Finally, we propose the AI-based joint Resource schEduling (ARE) framework, which includes two strategies, i.e., Deep neural network (DNN)-based and deep reinforcement learning (DRL)-based architectures. DNN-based solution with online incremental learning applies the global optimizer and therefore may have better performance than DRL-based architecture with online policy updating, but requires longer training time. Simulation results are also provided to verify the effectiveness of our proposed framework.


\section{AI driven H-MEC architecture}

\begin{figure*}[htpb]
	\centering
	\includegraphics[width=14cm]{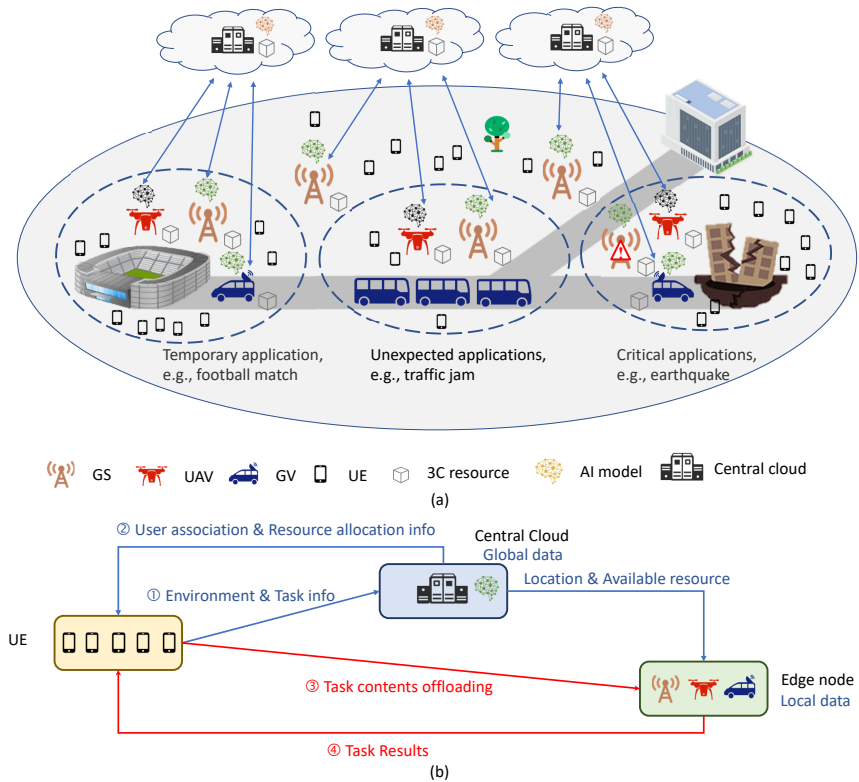}
	\caption{AI driven H-MEC system: (a) system architecture; (b) data flow.}
	\label{fig:fig2}
\end{figure*}

In this section, we show the AI driven H-MEC architecture in Fig. \ref{fig:fig2} (a), where the central cloud can not only provide 3C capacities for signal processing related tasks (e.g., fast Fourier transformation, encoding and decoding), but also provide computing resource for service related tasks (e.g., AI model training), due to its powerful accumulated processing capacities. 
In addition, we assume there are several distributed edge systems served by both fixed nodes (i.e., GSs) and mobile nodes (i.e., UAVs and GVs). Similar to the central cloud, the GSs, UAVs and GVs are all equipped with 3C resources. UAVs and GVs, due to their nature of flexibility, can be deployed swiftly on demand. In general, UAV moves much faster than the GV, but with less 3C resources. GV moves slower but holds more resource and also it normally has more energy on board, compared to UAV. GS has the most available resource but it is a fixed architecture. In addition, as UAVs can fly close to the user and therefore they can provide low-latency communication and services.  
Due to 3C resources available in H-MEC, AI based models can be trained in cloud platforms with different accuracy and abilities, depending on their available resource. Some complex model can be trained in the place with more resources, e.g., central cloud and then downloaded to the places with less resource, e.g., UAVs. Additionally, inference can be made in edge nodes, depending on the latency requirement of the applications or users. 
The features of different H-MEC components are summarised in Table \ref{tab:table1}.
\begin{table}[]
\centering\makegapedcells
\caption{Comparison for different H-MEC components.}
\label{tab:table1}
\begin{tabular}{|p{55pt}|p{35pt}|p{35pt}|p{35pt}|p{35pt}|}
	\hline 
	Components & Central Cloud & GS & GV & UAV\\ 
	\hline 
	3C Resource & Very Large & Large & Small & Very Small \\ 
	\hline 
	Mobility & None & None & Slow & Fast \\ 
	\hline 
	Serving Time & Unlimited & Unlimited & Long & Short \\ 		
	\hline 	
	Deployment  & Fixed & Fixed & 2D and restricted &  3D\\ 
	\hline   
	Response Time & Long & Long & Short & Short\\ 
	\hline 
	AI Model  & Very Complex & Complex & Simple & Simple  \\ 
	\hline
	AI Model Training  & Offline  & Incremental & Incremental & None  \\ 
	\hline
	Data Collection  & No & Yes & Yes &  Yes\\ 
	\hline 
	Data Type & Global  & Local & Local &  Local\\ 
	\hline 		
\end{tabular} 
\end{table}

Next, we give the data flow of one particular use case of the H-MEC, as shown in Fig. \ref{fig:fig2} (b), where the central cloud can collect the environment information (e.g., the number of UE and the channel state information) as well as task information (e.g., data size of the task and required computing resource of the task) from the UEs. Then, the central cloud can update the AI model, performing the online training and making decisions of the user association and resource allocation for each UE. Based on the decision received from the central cloud, each UE can offload the task to the nearby edge nodes, and then receive the results accordingly. Additionally, the central cloud also holds the information of the edge nodes (e.g., the available resource and location), in order to make the global resource allocation.

The proposed H-MEC architecture is particularly useful in the following scenarios:

(1) Temporary application: For instance, in a public event or a football match where there are a large number of people gathered, they may be interested in recording and exchanging high quality video contents. In these scenarios, it is very likely to have a large amount of traffic generated, particularly during the intervals of main events in the stadium.

(2) Unexpected application: For example, in the traffic jam, users inside the cars or buses would like to have data services using their mobile devices. Moreover, the traffic coordination centre may also need to communicate with the road units and cars so as to restore the traffic. This may create a large amount of data traffic, which may need the assistance of MECs.

(3) Critical application: For instance, in an emergency situation or natural disaster where an earthquake occurs, and people try to contact their relatives, which incurs a large amount of data traffic. Moreover, the rescue crews may need VR device to guide them to the place where needed. Those VR devices may need a large amount of computing resource. Therefore, the mobile edge nodes can be deployed to provide 3C resources.

\section{Research challenges}

Although H-MEC architecture has many benefits to be applied in dynamic scenarios as discussed before, the mobility of mobile nodes, e.g. UEs and MECs make the network topology highly unstable, which bring significant challenges as follows:

(1) Mobile edge node deployment: It is difficult to determine where to deploy the mobile edge nodes in dynamic environment, as the new edge servers joining the networks may lead to user offloading the tasks to them and then generate interference to the existing environment. Also, the mobile nodes, e.g., UAVs that are normally resource constrained may be difficult to meet the requirement of applications or users if lack of proper predictions. Dynamic programming \cite{5717933} can be used to calculate the optimal mobile edge node deployment. However, this method can only provide a snapshot of the optimal or sub-optimal solutions but fail to consider the correlation between different users in continuous time.

(2) Real-time decision making: The diverse requirements of different applications, time-varying content request, and the mobility of UEs make real-time decision a very challenging task. It is time-consuming for traditional convex optimization based methods, e.g., coordinate descent method \cite{8334188} to address this problem, as convex optimization based methods often require considerable number of iterations to reach a satisfactorily locally optimal solution. Moreover, convex optimization based methods may not be suitable for dynamic environment, as the optimization problem needs to be re-solved once the requirements or user patterns vary.

(3) Large-scale user associations: The typical use case of H-MEC, such as stadium or open air festival, may need to support massive UEs and applications. This problem normally includes integer variables and is NP-hard. 
Traditional solution was to apply the convex-based solutions or evolutionary algorithms. However, these solutions suffer from high complexity and are time-consuming. Moreover, branch and bound algorithm \cite{zhao2017energy} may be applied here but the search space of this method increases exponentially with the number of users and are computationally prohibitive.

(4) Resource management under specific constraints: In H-MEC, edge nodes can be served by mobile units, e.g., GVs and UAVs and they are normally resource-constrained. The coverage of mobile nodes, e.g., UAV may also be limited, as the communications links can be blocked by buildings. In addition, battery capacities could limit the capacity of mobile edge nodes as well. Therefore, all the above constraints need to be tackled jointly and properly, which create big challenges.   
Several AI based solutions, such as neural networks based methods are recently proposed by researchers but they are generally not suitable for dealing with the constraints.

(5) Caching deployment optimization: Caching has been identified as an important aspect by bringing storage functionality to edge servers. Deciding where/how/what to cache appropriate content have a profound effect on Quality of Service (QoS) requirement of UEs. Different from static MEC architecture, mobile edge nodes can be deployed on demand via tracking the mobility pattern of users and avoiding frequently updating the content from the core network. How to predict mobility patterns and content request information of users remains the main challenges.

(6) Security and privacy issues: This is critical for H-MEC systems, as the mobile edge nodes might not be able to detect an attack due to the lack of global information of the whole networks. Moreover, classical attack detection methods normally need manual feature engineering (e.g., feature design, selection and extraction) and therefore is hard to be implemented in dynamic environment. Thus, new approaches are highly required.

\section{AI-based solutions in H-MEC}
To address the above-mentioned challenges, in this section, we first discuss the AI-based solutions, which are well-known for its excellent modelling and prediction abilities. Then, we will give some tips in applying these solutions.

\begin{table*}[]
	\centering\makegapedcells
	\caption{Typical AI-based solutions in H-MEC.}
	\label{tab:table2}
	\begin{tabular}{|p{80pt}|p{60pt}|p{130pt}|p{170pt}|p{110pt}|}

		\hline
		AI-based solutions & Algorithm & Advantages & Disadvantages\\ \hline
		\multirow{2}{*}{Supervised learning} & GRU, bi-LSTM & Outstanding memory attribute and time series prediction ability. &  
		\begin{itemize}
			\item Labeled training data requirement.
			\item Unable to handle the constraint problems.
			\item Model is sophisticated and hard to deploy on the mobile equipment.
		\end{itemize} \\ \cline{2-3}  
		~   &CNN   & Mature technology and high recognition accuracy. &~\\ \hline
		
		\multirow{2}{*}{Unsupervised learning} & FCM & Soft clustering and no labeled sample requirement. & The solution is just an approximate version of the optimal result and  rely on initial data distribution. \\ \cline{2-3}
		& SAE & Feature learning automatically and no labeled sample requirement. & \\   \hline
		Reinforcement learning & DDPG & Learning from the environment and no labeled sample requirement. &The final results can be unstable and hard to reproduce \cite{henderson2018deep}. \\ \hline
		
	\end{tabular}
\end{table*}

\subsection{AI-based solutions}

(1) To deploy mobile edge nodes effectively and automatically, an unsupervised learning algorithm (e.g., the clustering algorithm) may be applied to analyse the locations, behaviours and preferences of UEs. Fuzzy C-Means (FCM) clustering \cite{bezdek1984fcm} is an improvement of common clustering algorithm, which adopts a soft fuzzy partition instead of the traditional rigid data classification, and thus could be applied to determine the dynamic deployment of mobile edge nodes. Another idea of deploying edge node could be to use the deep reinforcement learning (DRL) method (e.g., Deep Deterministic Policy Gradient, DDPG), which can learn optimal placement policies by considering the coverage, energy consumption and connectivity of edge nodes in the reward function, and place the mobile nodes intelligently \cite{liu2018energy}. 

(2) To address the real-time decision-making problem, Deep Neural Networks (DNN) could be applied as the real-time decision-making model due to the fact that once the training of DNN is completed, decisions can be made very fast by applying only a few simple algebra calculations. Moreover, by increasing the diversity of samples, DNN model is not sensitive to the dynamic environment. 
In addition, Recurrent Neural Networks (RNN) could be applied as well, due to its outstanding prediction and reasoning capabilities in real time. The Gated Recurrent Unit (GRU) network \cite{chung2014empirical}, which is a novel RNN, can make each recurrent unit to adaptively capture dependencies of different time scales. Also, the GRU simplifies the structure of RNNs by introducing reset and update gates, which can exploit the semantics and contexts from the input data (e.g., the varying channel quality information, CQI). For instance, H-MECs can apply the fast fading CQI to activate the reset gates and use the long-term large-scale channel fading information to activate the update gates and then make the real-time decision fully viable.

To solve the large-scale user association problems, Convolutional Neural Networks (CNNs) may be applied, due to its excellent feature extraction abilities. For instance, in H-MEC, CNN can be applied to identify important features (e.g., users' behaviours and channel quality) from the original large-scale information by applying several convolutional layers and then reduce the dimensionality of the original problem. To this end, the complexity of primal large-scale problem may be significantly reduced based on the extracted features. Another idea of solving the large-scale network optimization is to apply clustering algorithm, which divides original variables into several clusters. In this way, the original optimization problem can be divided into several small-scale sub-problems and tackled efficiently \cite{zhang2019toward}.

(4) For the resource management problems with several constraints, as mentioned before, it is difficult for the AI-based solutions to address them. This is because, AI-based solutions, e.g., neural networks, are normally designed for optimization without constraints and therefore the output of neural networks may not strictly satisfy all constraints. In this case, other methods, e.g., extra check may be conducted. Moreover, another layer may be attached to the networks dedicated to the feasibility check.

(5) For caching deployment in H-MEC, the bidirectional Long Short Term Memory (bi-LSTM) network \cite{sun2015voice} could be applied, as it can exploit both the previous and future contexts by analysing the data (e.g., video recoding clips) from two reverse directions. In particular, this scheme can deploy cache by considering the requested information at its previous and future states and predict the content request distribution. In the proposed H-MEC, bi-LSTM can be implemented to allow both fixed and mobile edge nodes to update their local content cache according to the mobility patterns of users while avoiding frequent access to the core network.

(6) For security and privacy issues in H-MEC, one may notice that the main obstacle of learning based methods lies in lack of samples. In other words, automatic feature selection and extraction are highly required. Auto-encoder algorithm may be applied. For example, H-MEC could benefit from a pre-training scheme of Stacked Auto-Encoder (SAE) \cite{9070170} for automatic feature learning. In particular, SAE can be applied to train an attack detection model with a mix of unlabelled normal/attack samples so that the model identifies patterns of attack and normal data by an auto-encoder scheme, this can in turn improve the accuracy of the attack detection model on unseen and mutated attacks. 

Moreover, in Table \ref{tab:table2}, we summarize the main advantages and disadvantages of different AI-based solutions in H-MEC.

\subsection{Tips}
In this subsection, we will give some tips on the design of deep learning (DL)-based solutions (e.g., CNN, GRU, bi-LSTM, SAE and DDPG), as the typical representatives of AI-based methods applied in H-MEC.

(1) Incremental learning: When applying DNN in H-MEC, one may need to continually train and adjust the parameters in response to the fast changing environment. Incremental learning could be applied here to update the DNN model dynamically \cite{Zhou_2017_CVPR}. In H-MEC, edge servers can track the variations of the environment and update the training data periodically and applied to re-train the learning model to guarantee that the model performs well even when the environment is constantly changing.

(2) Compressing learning: As the mobile edge nodes are resource/energy constrained, it is important to reduce the energy and computation consumption during the leaning process. Therefore, the compressing learning could be applied. Compressing learning can reduce parameters of DNNs while mitigating result accuracy loss. Since DNNs are usually extremely over parameterized, they are capable of compression. Several approaches have been proposed to facilitate this process, such as network pruning, knowledge distillation, weight quantification and lossless compression \cite{xie2019energy}. 

(3) Experience Learning: In H-MEC, it is important to get the high-quality data as training samples, otherwise, the trained model may be biased. To achieve this goal, experience learning may be applied here to find the optimal solutions from the historical experience data \cite{zhang2019toward}. 

\section{AI-based joint Resource schEduling (ARE) framework}

In this section, we will introduce the AI-based joint Resource schEduling (ARE) framework with two different strategies, (i.e., DNN- and DRL-based architectures) to show the potential of AI-based solutions in H-MEC when applying it to the dynamic environment.

\subsection{System Model and Problem Formulation}
\subsubsection{System model}
We consider the MEC system with multiple UEs and edge nodes consisting of UAVs, GVs and GSs, as shown in Fig. \ref{fig:fig2}. We assume that each UE has one computation task, which can be executed either locally or by one of the edge nodes. We model the computational-intensive task of the $i$-UE as $U_i = (F_i, D_i)$, $\forall i \in \mathcal{N}$, where $F_i$ denotes the computing resource required by the task and $D_i$ denotes the data size of the task if offloading is decided.
\subsubsection{Computing model}
The local task execution time for the $i$-th UE is determined via $F_i$ divided by the computation capacity of the $i$-th UE (in CPU cycles per second). 
Also, the edge task execution time for the $i$-th UE is determined via $F_i$ divided by the allocated computing resource from the edge nodes. 
\subsubsection{Communication model}
For the remote execution in edge nodes like UAV, GV and GS, we assume that the UE offloads its task via orthogonal frequency division multiplexing (OFDM) channels, which means that there is no interference between each other. The communication delay for the $i$-th UE is determined via $D_i$ divided by the transmission data rate between UE and MEC (e.g., UAV, GV or GS).
\subsubsection{Problem formulation}
We aim to obtain an online algorithm to minimize the sum of weighted latency for all the tasks, by jointly optimizing the user association and resource allocation in real time, while considering dynamic environment, i.e., the number and the locations of UEs may vary. One can formulate the optimization problem as follows: 
\begin{itemize}
	\item Objective function: the minimization of total weighted task latency of all the UEs (i.e., the summation of communication delay and remote task execution delay if offloading or the local execution time if completing the tasks locally);
	\item  Decision variables: user association, resource allocation, and the locations of GVs and UAVs;
	\item Constraint $C1$: tasks can be executed either locally or by one of the edge nodes;	
	\item Constraint $C2$: the coverage of each UAV;
	\item Constraint $C3$: the computing resource available in each node (i.e., UEs, GSs, GVs and UAVs).
\end{itemize}  

One can see that the above problem is a mixed integer non-linear programming (MINLP), which is challenging to address. This is because the user association is binary whereas the resource allocation and the location variables of UAVs and GVs are continuous. 
This problem becomes even more challenging if we consider the dynamic environment (which means the system parameters, i.e., the channel state information, the number and the locations of UEs may change). 

We decompose the formulated problem into two sub-problems: 1) deployment problem of mobile edge nodes; and 2) resource scheduling and decision-making problem. We first apply a clustering algorithm to locate UAVs and GVs\cite{8907406}, and then we apply DNN-based and DRL-based ARE framework to conduct the decision making and resource allocation for each UE. The core part of the ARE framework is the application of DNN to make the real-time decision. Different from current AI-based contributions, we  input the state information of one UE to the DNN at a time. This modification has the following benefits: (1) The input dimension of the DNN only depends on the number of H-MECs and is not related to the number of UEs. In general, the number of access points or edge nodes changes much slower than the number of UEs accessing the network. Hence, our DNN can be applied in a long time once it is trained, which is more practical in the real-world environment. (2) Since each time, we only input the information of one user, we can reduce the dimension of input data and then reduce the training complexity, which is suitable for large-scale networks with large number of UEs. Next, we adopts the epoch register to collect the input and output information of all the UEs at this epoch for the sample optimization and DNN training, which can train the DNN by considering the information of all the UEs. Finally, we introduce two different online learning mechanisms to train the DNN for tracking the variations of the real-world scenarios. Next, we will introduce ARE framework in more details. 

\subsection{DNN-based ARE framework}
\begin{figure*}[htpb]
	\centering
	\includegraphics[width=18cm]{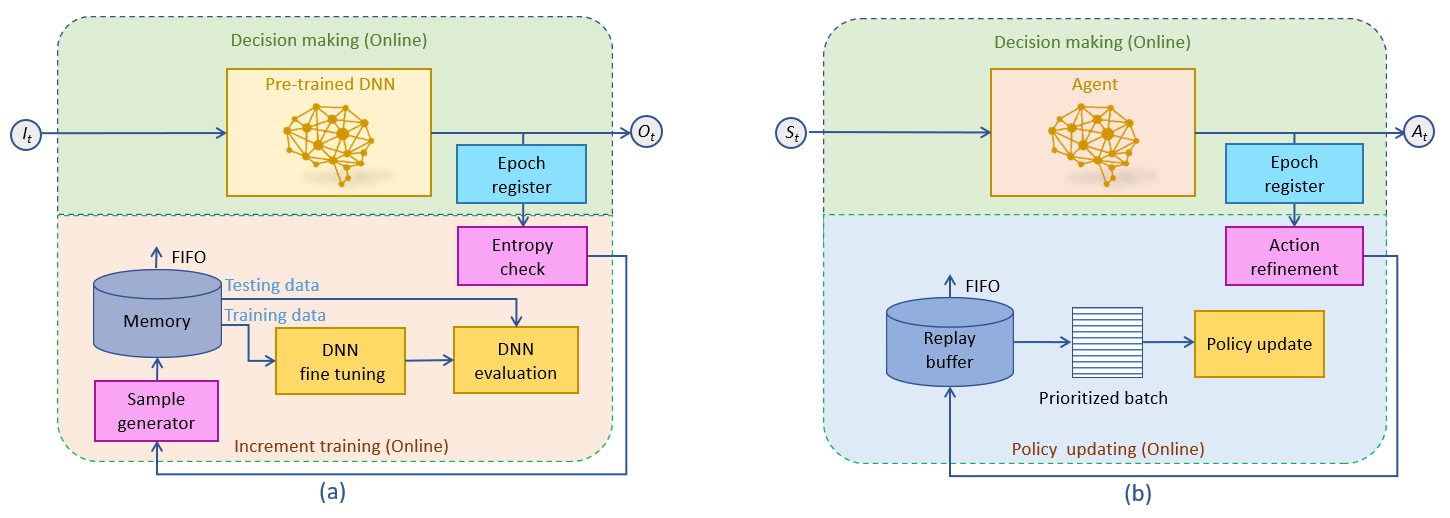}
	\caption{The proposed ARE framework.\\ (a) DNN-based architecture; (b) DRL-based architecture.}
	\label{fig:fig5}
\end{figure*}

We first present the DNN-based ARE framework, as shown in Fig. \ref{fig:fig5} (a), which includes the offline pre-training stage, online decision making process and the incremental learning stage. Note that the parameter tuple $(\mathcal{H}_i, F_i, D_i, \mathcal{W})$ of the $i$-th UE is applied as the input to the DNN, and the decision tuple $(a_i,f_i)$ is used as the output of the DNN, where $\mathcal{H}_i=\left\{h_{ij},  \forall j \in \mathcal{M} \right\}$ is a set of the channel gains $h_{ij}$ between the $i$-th UE and the $j$-th edge nodes. $\mathcal{W}=\left\{w_j, \forall j \in \mathcal{M} \right\}$ includes the average allocated computing resources for each UE of the edge nodes, which can be counted from the historical data. In the following, we describe each stage of the DNN-based framework.

\subsubsection{Offline pre-training stage}
The training phase can be carried out in the cloud as it holds a large amount of computing resource. Firstly, lots of historical data is collected. To this end, the sample generator is applied for solving the original problem. In general, the optimization algorithm can be divided into the following three categories: (1) The exhaustive search is applied to obtain the sample if the search space is small; (2) The mixed-integer programming solvers (e.g., CPLEX) can be applied when the search space is medium; (3) The global heuristic algorithms (e.g., Genetic Algorithm or Particle Swarm Optimization) may be adopted to obtain samples when the search space is large. The sample generator is carried out repeatedly until sufficient samples are collected. Then, the supervised learning is applied to train the DNN until the evaluation conditions are satisfactory. 
 \subsubsection{Online decision stage}
The pre-trained DNN can be implemented for online decision making process. To do this, at epoch $t$, the system information $\mathcal{I}_{i,t}=\left\{\mathcal{H}_{i,t}, F_{i,t}, D_{i,t}, \mathcal{W}_{t}\right\}$ of the $i$-th UE is input to the DNN and obtain the corresponding solutions $\mathcal{O}_{i,t}=\left\{a_{i,t},f_{i,t}\right\}$  of the $i$-th UE, with only some simple algebraic calculations instead of solving the original optimization problem. An epoch register is applied to store solutions $\left\{\mathcal{I}_{i,t},\mathcal{O}_{i,t}\right\}, \forall i \in \mathcal{N}$ of all the UEs at epoch $t$ for considering the system information of all UEs in the incremental learning.

\subsubsection{Online incremental learning stage}
Incremental learning is applied for tracking the variations of the dynamic environment, which can improve the proposed DNN through continuous fine-tuning rather than repeatedly re-training the network from scratch. The procedure of incremental learning is described as follows: Firstly, an average entropy check is applied to inspect the output of DNN at epoch $t$, which can decide if the collection of new samples are needed. Outputs with higher entropy are expected to contribute more to elevate the current DNN’s performance. Therefore, entropy check is applied by applying a simple threshold evaluation. If the average output entropy of all the UEs at current epoch is larger than the threshold, the current state information is sent to the sample generator for global optimization and the new samples are generated to store in the memory. The memory is the dynamic database with fixed-size, and the first-in first-out (FIFO) scheduling policy is applied when the memory is full. In our framework, the memory is applied as the sample database to fine-tune the DNN, which means the negative error gradient of the current iteration is added to the weights of the DNN fine-tuned in the previous iteration, in real-time processes.



\subsection{DRL-based ARE framework}
 
In this subsection, we introduce the second type of ARE framework, which is the DRL-based architecture, as shown in Fig. \ref{fig:fig5} (b), which includes online decision making stage and policy updating stage. Note that we apply the DNN as the agent of DRL. The state of the $i$-th UE is given by $\mathcal{S}_{i,t}=\left\{\mathcal{H}_{i,t}, F_{i,t}, D_{i,t}, \mathcal{W}_{t}\right\}$ and the action of the $i$-th UE is given as $\mathcal{A}_{i,t}=\left\{a_{i,t},f_{i,t}\right\}$, and the reward is the reciprocal of our objective function. In the following, we describe each stage of the proposed framework in details.

\subsubsection{Offloading decision making stage} 

At the epoch $t$, the agent whose parameters are represented as the offloading policy $\pi_{t}$, which can be deployed for generating online scheduling decision $\mathcal{A}_{i,t}$ according to the state $\mathcal{S}_{i,t}$. Then the epoch register is applied to collect $\left\{\mathcal{S}_{i,t},\mathcal{A}_{i,t}\right\}, \forall i \in \mathcal{N}$ of all UEs at this epoch
for considering the states of all UEs in the action refinement. As the decision making from the offloading policy $\pi_{t}$ has low computing complexity via forward networks of DNN, the decision $\mathcal{A}_{i,t}$ can be output in real time. 

\subsubsection{Offloading policy updating stage} 
The action refinement is applied as an efficient exploration to find sub-optimal action $\mathcal{A}_{i,t}^{*}$ compared to the traditional random search process (e.g., $\epsilon$-greedy). In general, the action refinement can be divided into two categories: (1) The local exhaustive search is applied (e.g., K-Nearest Neighbour) when the action space is small; (2) The local heuristic search method is applied (e.g., Simulated Annealing or Tabu Search) when the action space is large. The improved $\left\{\mathcal{S}_{i,t},\mathcal{A}_{i,t}^{*} \right\}$ explored by the action refinement is selected as the new transition and appended to the replay buffer. Then, a batch of transitions are drawn from the buffer by prioritized experience replay, and the agent is trained and the offloading policy is updated from $\pi_{t}$ to $\pi_{t+1}$. The new offloading policy $\pi_{t+1}$ is applied in the epoch $t+1$ to generate the offloading decision $a_{t+1}$ according to the new $s_{t+1}$. 

Moreover, the above two stages are alternately performed and therefore the offloading policy can be gradually improved in the iteration process. 

\subsection{Summary of above DNN- and DRL-based architecture} 
In this section, we will provide the summary and comparison of the above DNN- and DRL-based ARE architecture.
For the DNN-based framework, we employ incremental learning to track the variations of the dynamic environment. In addition, a novel entropy check is applied to inspect the output of DNN and collect new system information from the varying environment. As we use the global optimizer as the sample generator, the computing time of online training is longer than the DRL-based ARE framework, but the performance is better than the DRL-based framework, especially for large-scale MEC systems with slowly changing environment.

In the DRL-based ARE framework, we adopt DRL to update offloading policy dynamically for the varying environment. An extra action refinement is introduced to explore actions for improving the efficiency and robustness of the DRL-based framework. For achieving online policy updating, we adopt the local optimizer as the action refinement tool, and therefore the computing time is shorter than the DNN-based framework, but the performance decreases when the action space increases. This framework is suitable for small to medium-scale MEC systems with fast changing environment.

\subsection{Simulation Results}

In the simulation, we assume that there are 50 UEs, 2 UAVs, 1 GV and 1 GS in a 50m $\times$ 50m squared zone, with the coordinate of GS as (25m, 25m). We apply the line 3$x$+2$y$-180=0 to depict the position of road for GV. The bandwidth is set as 1 MHz, the computational capability of UE, UAV, GV and GS is set to $10^{9}$ cycles/s, 15$\times$$10^{9}$ cycles/s, 30$\times$$10^{9}$ cycles/s and 50$\times$$10^{9}$ cycles/s, respectively. Each user is randomly distributed with the maximal velocity of 1m/s. The environment data (e.g., $\mathcal{H}_{i,t}, F_{i,t}, D_{i,t}, \mathcal{W}_{t}$) is collected every 3 seconds for the training of ARE. For the DNN-based framework, the DNN structure includes an input layer, two hidden layer (64 and 32 neurons) and an output layer. The learning rate and iteration number of DNN are set to 1.5 and 500, respectively. For the DRL-based framework, the agent employs the same structure of DNN; the replay buffer size is set to 10000; and the batch size is set to 1000. The Simulated Annealing is applied as the action refinement algorithm. All simulations for the DNN and DRL are carried out in Matlab 2020 environment running on a i7-6500U CPU with 8GB RAM and 512G SSD. The DNN and DRL are implemented by the deep learning toolbox. The total simulation time of the DNN-based ARE and DRL-based ARE are 123.56 seconds and 43.32 seconds, respectively.

\begin{figure}
	
	\centering
	\subfigure[]{		
		\includegraphics[width=9cm]{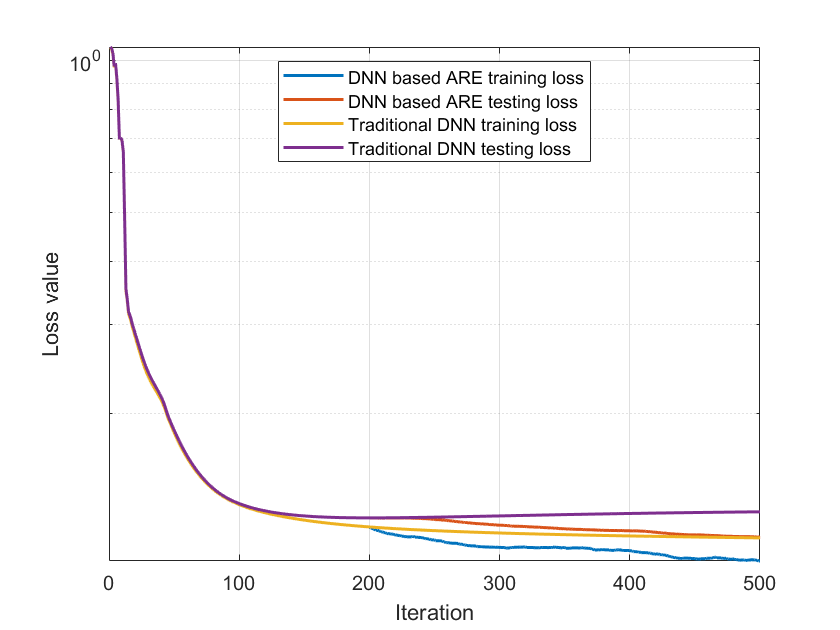}
		\label{fig7:a}
	}
	\subfigure[]{	
		\centering		
		\includegraphics[width=9cm]{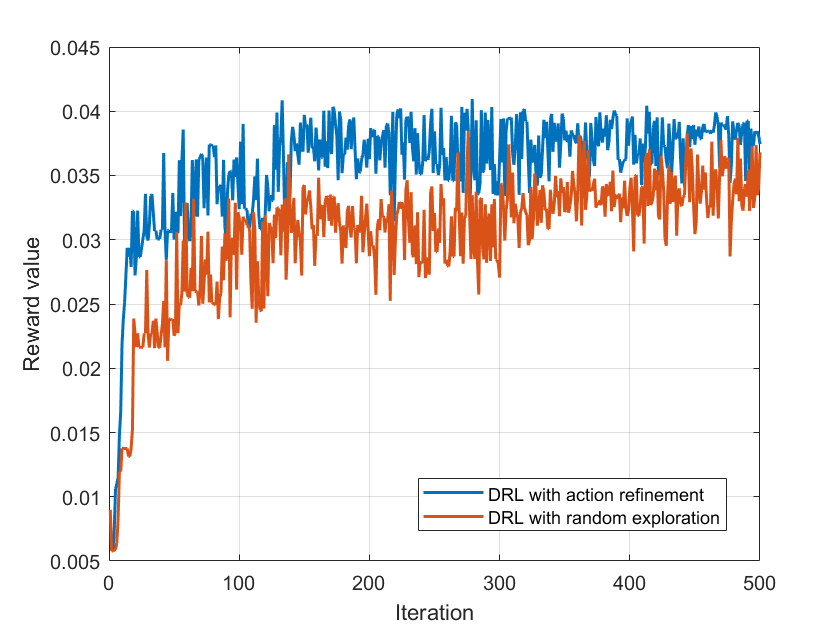}
		\label{fig7:b}
	}
	\caption{Performance comparison: (a) Training and testing losses of DNN-based ARE; (b) Reward values of DRL-based ARE.}
	\label{fig:fig7}
\end{figure}

In Fig. \ref{fig:fig7} (a), we compare the performance of the DNN-based ARE model  and the traditional DNN model without incremental learning. It can be seen that the DNN-based ARE model achieves lower training and testing losses than the traditional DNN. This is because the incremental learning can enhance the learning ability of DNN in dynamic environment and allow the DNN to learn latest information continuously. We can also see that the difference between the testing loss and the training loss of traditional DNN increases gradually without incremental learning when the environment is changing.

Fig. \ref{fig:fig7} (b) further characterizes the benefits of the proposed DRL-based model with the action refinement by using the performance metric, i.e., reward during the online policy updating stage. One can see that the proposed model with action refinement not only converges to a higher reward than traditional DRL but also achieves faster convergence speed. This is because the action refinement is an efficient exploration process to find sub-optimal action compared to the traditional random search process.

Then, we evaluate the performance of DNN- and DRL-based ARE framework by comparing with the following offloading schemes:
\begin{itemize}	
	\item Random offloading (Random) denotes that the offloading decision is decided randomly for each UE.   
	
	\item Greedy offloading (Greedy) denotes that all the UEs offload the tasks to the nearest edge nodes. 
	
	\item Local execution (Local) denotes that all UEs decide to execute the task locally.		
\end{itemize}

\begin{figure}[htpb]
	\centering
	\includegraphics[width=9cm]{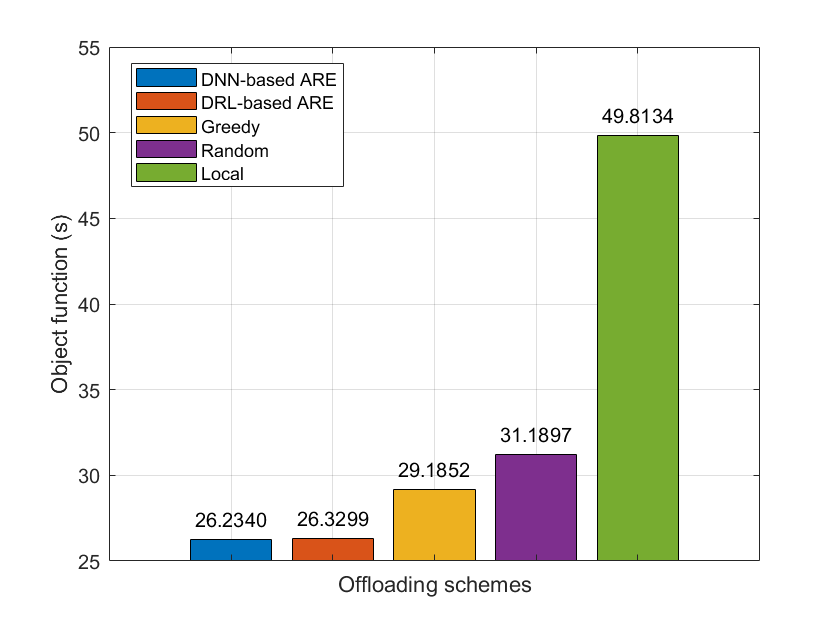}
	\caption{Performance comparison of the objective function between different offloading schemes..}
	\label{fig8}
\end{figure}

Fig. \ref{fig8} shows the total task latency versus different offloading schemes. One can see that compared to Greedy, Random and Local offloading schemes, DRL-based ARE achieves similar performance to the DNN-based ARE but performs much better than the other algorithms.

\section{Conclusions}

In this paper, we have studied the AI driven H-MEC architecture, which is expected to be applied in dynamic environment. We have discussed the key challenges of the H-MEC architecture and the possible AI-based solutions. Moreover, we provide an ARE framework with two different strategies. In the DNN-based ARE framework, online incremental learning stage is applied for tracking the dynamic environment, while in the DRL-based ARE framework, online policy updating is used to adjust the policy of DRL. Moreover, we only input the state information of one UE to the ARE framework each time, which is more practical for the scenarios with changing number of UEs. Simulation results have been provided to show the effectiveness of the proposed framework.

The future research directions can be summarized in the following: (1) we will consider to test the proposed framework in the real-world test-bed or apply the real data set from the mobile operators; (2) we plan to predict the mobility pattern and requests of the mobile users, in order to improve the performance of the whole networks; (3) we will further enhance the security and privacy of the ARE framework.


%

%




\bibliographystyle{ieeetran}
\bibliography{bare_jrnl}

\end{document}